\documentstyle{mn}                
\input{epsf}
\begin{document}
%version 29.01.97
    
\def\ee{$e^\pm$}
\def\g{$\gamma$}
\def\nh{N_{\rm H}}
\def\af{A_{\rm Fe}}
\def\taut{\tau_{\rm T}}
\def\ginga{{\it Ginga}}
\def\asca{{\it ASCA}}
\def\heao{{\it HEAO-1}}
\def\ec{E_{\rm c}}
\def\efe{E_{\rm Fe}}
\def\sfe{\sigma_{\rm Fe}}
\def\ife{I_{\rm Fe}}
\def\loc{\ell_{\rm l}}

% Next lines define "less than or approximately
% equal to", "greater than or approximately equal to", and "approximately
% proportional
\newbox\grsign \setbox\grsign=\hbox{$>$} \newdimen\grdimen %
\grdimen=\ht\grsign
\newbox\simlessbox \newbox\simgreatbox \newbox\simpropbox
\setbox\simgreatbox=\hbox{\raise.5ex\hbox{$>$}\llap
     {\lower.5ex\hbox{$\sim$}}}\ht1=\grdimen\dp1=0pt
\setbox\simlessbox=\hbox{\raise.5ex\hbox{$<$}\llap
     {\lower.5ex\hbox{$\sim$}}}\ht2=\grdimen\dp2=0pt
\setbox\simpropbox=\hbox{\raise.5ex\hbox{$\propto$}\llap
     {\lower.5ex\hbox{$\sim$}}}\ht2=\grdimen\dp2=0pt
\def\simgreat{\mathrel{\copy\simgreatbox}}
\def\simless{\mathrel{\copy\simlessbox}}

\topmargin = -1cm

\title[X-ray and gamma-ray observations of Cyg X-1]
{Simultaneous X-ray and gamma-ray observations of\\
Cyg X-1 in the hard state by Ginga and OSSE} 

\author[M. Gierli\'nski et al.]
{\parbox[]{6.8in} {Marek Gierli\'nski$^{1,2}$, Andrzej A. Zdziarski$^{2,3}$, 
Chris Done$^4$, W. Neil Johnson$^5$, 
Ken Ebisawa$^3$, Yoshihiro Ueda$^6$, Francesco Haardt$^7$, and Bernard F. 
Phlips$^{8,5}$}\\ 
 $^1$Astronomical Observatory, Jagiellonian University, Orla 171, 30-244 
Cracow, Poland \\ 
$^2$N. Copernicus Astronomical Center, Bartycka 18, 00-716 Warsaw, Poland \\ 
$^3$Laboratory for High Energy Astrophysics,
NASA/Goddard Space Flight Center, Greenbelt, MD 20771, USA \\
$^4$Department of Physics, University of Durham, Durham DH1 3LE, UK \\
$^5$E. O. Hulburt Center for Space Research,
Naval Research Laboratory, Washington, DC 20375, USA \\
$^6$Institute of Space \& Astronautical Science, 3-1-1, Yoshinodai, 
Sagamihara-shi, Kanagawa 229, Japan \\
$^7$Department of Astronomy \& Astrophysics, Gothenburg University, 41296 
Gothenburg, Sweden\\
$^8$Universities Space Research Association, Washington DC 20024, USA}

\date{Accepted 1997 February 10, Received 1996 October 15}

\maketitle

\begin{abstract}
We present four X-ray/\g-ray spectra of Cyg X-1 observed in the hard (`low') 
state simultaneously by \ginga\/ and {\it GRO\/} OSSE on 1991 June 6. The four 
spectra have almost identical spectral form but vary in the normalisation 
within a factor of two. The 3--30 keV \ginga\/ spectra are well represented by 
power laws with an energy spectral index of $\alpha\sim 0.6$ and a Compton 
reflection component including a fluorescent Fe K$\alpha$ corresponding to the 
solid angle of the reflector of $\sim 0.3\times 2\pi$. These spectra join 
smoothly on to the OSSE range ($\geq 50$ keV) and are then cut off above $\sim 
150$ keV. The overall spectra can be modelled by repeated Compton scattering 
in a mildly-relativistic, thermal plasma with the optical depth of $\tau\sim 
1$. However, the high-energy cutoff is steeper than that due to 
single-temperature thermal Comptonisation.  It can be described by a 
superposition of dominant $\tau\sim 1$--2, thermal emission at $kT\sim 100$ 
keV and a Wien-like component from an optically-thick plasma at $kT\sim 50$ 
keV. 

The X-ray spectra do not show the presence of an anisotropy break required if 
thermal Compton scattering takes place in a corona above a cold disc. Also, 
the flat spectral index shows that the plasma is soft-photon starved, i.e., 
the luminosity in incident soft X-ray seed photons is very much less than that 
in the hard X-rays. Furthermore, the observed solid angle of the reflector is 
significantly less than $2\pi$. These facts taken together strongly rule out a 
disc-corona geometry. Rather, the observed spectra are consistent with a 
geometry in which the cold accretion disc (which both supplies the seed soft 
X--rays and reflects hard X-rays) only exists at large radii, while the 
Comptonising hot plasma is located in an inner region with no cold disc. This 
hot plasma consists of either pure \ee\ pairs if the source size is
$\sim 5$ Schwarzschild radii or it contains also protons if the size is 
larger. 

\end{abstract}


\section{INTRODUCTION}
\label{s:intro}

Cyg X-1, one of the brightest X-ray sources in the sky, was discovered 
in X-rays more than 30 years ago (Bowyer et al.\ 1965). This binary 
system is at a distance of $\sim 2.5$ kpc and consists of the O9.7 Iab 
type supergiant HDE 226868 (Gies \& Bolton 1986) orbiting around a 
compact object with a period of 5.6 days. The mass of the unseen 
companion is significantly larger then 5 $M_\odot$ (Dolan 1992), 
strongly suggesting that it is a black hole. Focused wind accretion 
(Gies \& Bolton 1986) from the primary star (which is extremely close to 
filling its Roche lobe) drives the powerful source of the X-ray 
radiation.  The X-ray emission of Cyg X-1 exhibits strong variability on 
all time scales from milliseconds to months (e.g. Miyamoto \& Kitamoto 
1989; Priedhorsky, Terrel \& Holt 1983). However, it spends most of the 
time in the hard (so-called `low') X-ray state, characterised by a low 
flux in soft X-rays and strong hard X-ray and soft \g-ray fluxes (e.g., 
Phlips et al.\ 1996, hereafter P96). In this state, the 50 keV flux 
changes by about an order of magnitude but the spectral variability is 
modest (P96). From time to time, Cyg X-1 is in the soft (`high') state 
(e.g., Liang \& Nolan 1984, Cui et al.\ 1997), with a strong soft X-ray 
emission and a relatively weak tail in hard X-rays.

The hard-state spectrum of Cyg X-1 at $\simgreat 3$ keV consists 
typically of a power law with an energy spectral index of $\alpha\sim 
0.6$--0.7 and a Compton-reflection continuum component, prominent above 
10 keV (Done et al.\ 1992; Ebisawa et al.\ 1996, hereafter E96). This 
component also includes an Fe K edge above $\sim 7$ keV and an Fe K$\alpha$ 
fluorescence line (Barr, White \& Page 1985; Marshall et al.\ 1993; E96). The 
results of BBXRT (Marshall et al.\ 1993) and \asca\/ (E96) indicate that the 
line is narrow, with the width of $\sigma\simless 0.2$ keV, suggesting its 
origin in outer parts of an accretion disc. The edge energy suggests the 
reflecting surface is mildly ionised, with Fe XI as the most abundant Fe 
species (E96). 

The soft \g-ray spectra of Cyg X-1 in the hard state are steeply cut off 
above $\sim 150$ keV (P96). P96 obtained good fits to the {\it GRO\/} 
OSSE spectra of Cyg X-1 with exponentially cut off power laws. However, 
the X-ray spectral index implied by those fits is much harder than that 
observed in X-rays, showing that the broad-band X-ray/\g-ray (hereafter 
X\g) spectrum is more complex. 

The most likely physical process responsible for the intrinsic continuum 
appears to be thermal Comptonisation of soft X-rays from thermal 
emission of a cold accretion disc at a temperature of $\sim 100$ eV 
(e.g., Ba{\l}uci\'nska-Church et al.\ 1995). Haardt et al.\ (1993) 
pointed out that Comptonisation may take place in an optically 
thin plasma rather than in an optically thick one (as fitted by, e.g., 
Sunyaev \& Tr\"umper 1979). 

In this work, we present four 2--1000 keV spectra of Cyg X-1 observed in 
the hard state simultaneously by \ginga\/ and OSSE on 1991 June 6. This 
provides us with an unprecedented opportunity to study physical 
processes taking place in the X\g\ source as well as the source 
geometry. We consider thermal and nonthermal Comptonisation, Compton 
reflection, \ee\ pair production, and constraints on the sizes and 
relative covering factors of the X\g\ source and cold matter. 

\section{THE DATA}
\label{s:data}

Cyg X-1 was observed four times by \ginga\/ on 1991 June 6 during the OSSE 
viewing period 2 (P96). Overlapping observation periods provide us with four 
sets of nearly simultaneous data. The orbital phase during the four 
observations was from 0.70 to 0.85. Table 1 gives the log of observations. We 
adopt 4661 cm$^2$ as the \ginga\/ effective area, which is the standard value 
used in the Leicester data base (D. Smith, private communication). The high 
voltage of the \ginga\/ LAC detector (Turner et al.\ 1989) was reduced in this 
observation, which lead to a usable energy range of 2--30 keV (E96). A 1 per 
cent systematic error was included in each \ginga\/ channel. 

\begin{table*}
\label{t:log}
\centering
\caption
{The log of observations in UT on 1991 June 6. The counts of \ginga\/ and OSSE 
are for 2--30 keV and 50--150 keV, respectively.} 
 \begin{tabular}{lcccccccc}
\hline
&\ginga && &&OSSE&&\\ 
Data set & Start & End & Live time [s] &Counts/s
&Start & End & Live time [s] &Counts/s\\
1 &$00^{h}17^{m}50^{s}$ &$02^{h}09^{m}50^{s}$  &2304 &$3410\pm 10$
&$00^{h}02^{m}32^{s}$ &$00^{h}37^{m}29^{s}$ &1806 &$455\pm 4$\\
& & & & &$01^{h}23^{m}21^{s}$ &$02^{h}11^{m}09^{s}$ &2315 &$554\pm 4$\\ 
2 &$04^{h}43^{m}26^{s}$ &$06^{h}29^{m}02^{s}$  &888&$3890\pm 10$
&$04^{h}28^{m}46^{s}$ &$05^{h}03^{m}43^{s}$ &1741 &$456\pm 5$\\
& & & & &$06^{h}02^{m}26^{s}$ & $06^{h}50^{m}29^{s}$ &2299 &$665\pm 5$\\
3 &$11^{h}02^{m}38^{s}$ &$14^{h}24^{m}46^{s}$  &2828&$2290\pm 10$
&$10^{h}42^{m}50^{s}$ &$11^{h}15^{m}36^{s}$ &1708 &$400\pm 4$\\
& & & & &$12^{h}14^{m}47^{s}$ &$12^{h}54^{m}07^{s}$ &2010 &$371\pm 4$\\ 
& & & & &$13^{h}48^{m}41^{s}$ &$14^{h}32^{m}23^{s}$ &2257 &$307\pm 4$\\ 
4 &$20^{h}22^{m}14^{s}$ &$20^{h}43^{m}58^{s}$  &1272&$2080\pm 10$
&$20^{h}02^{m}15^{s}$ &$20^{h}32^{m}50^{s}$ &1629 &$316\pm 4$\\ 
\hline
\end{tabular}
\end{table*}
 
The OSSE data are from 50 keV to 1000 keV. They include estimated 
systematic errors computed from the uncertainties in the low energy 
calibration and response of the detectors using both in-orbit and 
prelaunch calibration data. The energy-dependent systematic errors are 
expressed as an uncertainty in the effective area in the OSSE response. 
These systematic errors were added in quadrature to the statistical 
errors. The former are most important at the lowest energies 
(approximately 3\% uncertainty in effective area at 50 keV, decreasing 
to 0.3\% at 150 keV and above). 

Figure 1 shows the four unfolded spectra corresponding to the four data 
sets. (The solid curves represent a two-component model described in 
Section 3.1 below.) The ratio of the highest 2--1000 keV model 
flux to the lowest one, obtained during the second and the fourth 
observation, respectively (separated by about 15 hours), is $\sim 1.8$.  

\begin{figure}
\begin{center}
\leavevmode
\epsfxsize=8.4cm \epsfbox{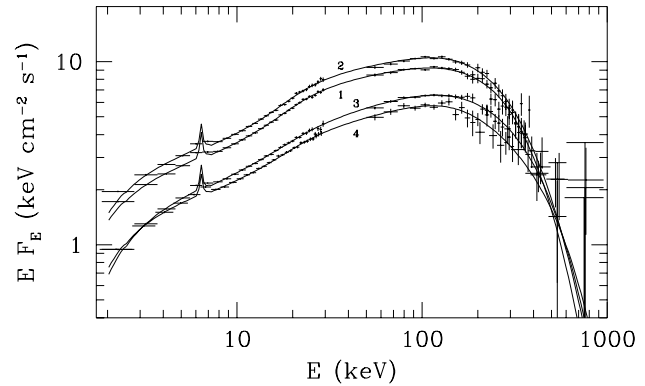}
\end{center}
\label{fig:spec}
\caption{
The four spectra of Cyg X-1, labelled by the observation number (see Table 1), 
observed simultaneously by \ginga\/ and OSSE on 1991 June 6. The solid curves 
represent the two-temperature model described in Section 3.1. The data have 
been rebinned for clarity of the display. } 
\end{figure}

\section{RESULTS}
\label{s:res}

\subsection{Spectral fits}

In our fits, we use the {\sc xspec} spectral fitting package version 9 (Arnaud 
1996). Our model spectra are absorbed by a column density, $\nh$, which we 
constrain to be $\geq$ the Galactic column, $6 \pm 2\times 10^{21}$ cm$^{-2}$ 
(Ba{\l}uci\'nska-Church et al. 1995). We obtain absorption larger than the 
Galactic one for some models, especially for the data sets 3 and 4 (Tables 2 
and 3), which can be attributed to transient dips associated with the 
intervening gas (Kitamoto et al.\ 1984). In X-rays, the spectra consist 
clearly of two components: an underlying power law and a component due to 
reflection from the surface of an accretion disc including both continuum 
Compton reflection and a fluorescent Fe K$\alpha$ line (e.g., George \& Fabian 
1991). 

We first fit the \ginga\/ data only with the incident continuum in the 
form of a power law e-folded with $\ec=300$ keV together with the 
Compton reflection component and an iron line. In the spectral fitting 
we ignore the data below 3 keV due to the presence of a soft excess at 
lower energies (E96), but we show these data in the plots (Figs.\ 1--4). 
We use angle-dependent Compton reflection Green's functions of Magdziarz 
\& Zdziarski (1995).  The disc inclination is assumed to be less than 
$67^\circ$ (Dolan \& Tapia 1989), and in most fits the most probably 
inclination of $30^\circ$ is used (Gies \& Bolton 1986). The luminosity 
intercepted by the reflecting medium equals $\Omega/2\pi$ times the 
luminosity emitted outward by the X\g\ source, where $\Omega$ is the 
solid angle subtended by the reflector as seen from the source of 
isotropic radiation. In some models, the reflecting medium is allowed to 
be ionised with an ionisation parameter, $\xi=L/(nr^2)$, where $L$ is 
the 5 eV--20 keV luminosity in a power law spectrum and $n$ is the 
density of the reflector located at distance $r$ from the illuminating 
source (Done et al.\ 1992). The reflector temperature is kept at 130 eV, 
which corresponds to the temperature of the soft X-ray spectral 
component (Ba\l uci\'nska-Church et al.\ 1995). The abundances are from 
Anders \& Ebihara (1982) except that the relative Fe abundance, $\af$, 
is a free parameter. The ion edge energies and opacities are from 
Reilman \& Manson (1979) except for the Fe K-edge energies taken from 
Kaastra \& Mewe (1993). 

We find little intrinsic spectral variability in the \ginga\/ data, so we use 
the average \ginga\/ spectrum to better constrain the parameters in X-rays. We 
obtain $\alpha=0.59^{+0.03}_{-0.03}$, $\Omega/2\pi = 0.34^{+0.05}_{-0.05}$, 
$\af=2.1^{+0.3}_{-0.3}$ ($\chi^2=21/22$ d.o.f.). The ionisation of the 
reflector is weak, $\xi=23^{+26}_{-22}$ erg cm s$^{-1}$. There is a 
distinct Fe K$\alpha$ line in the spectrum, which we model here as a Gaussian 
centred at 6.4 keV and with the width of 0.1 keV, consistent with both the 
\ginga\/ resolution not allowing us to resolve the line and the \asca\/ 
results indicating the line is narrow (E96). The line equivalent width is 
EW=$74^{+26}_{-22}$ eV. These results are similar to those obtained for other 
\ginga\/ observations (e.g.\ E96). 

In the fits with both \ginga\/ and OSSE below, we examine the effect of 
allowing the relative normalisation of the two data sets to be free within 
$\pm 15$ per cent (which is a conservative limit to the relative calibration 
uncertainty of the two instruments). However, we find the relative 
normalisation to be consistent with unity as allowing it to be free leads to 
marginal fit improvements only. Therefore, in tables and figures below we use 
the actual normalisation of the \ginga\/ and OSSE spectra. 

When we extrapolate the \ginga\/ model to the OSSE range, there is a good 
match around 50 keV, but the OSSE data require a cutoff in the spectrum at 
$\simgreat 150$ keV. However, a simple e-folded power law together with its 
reflection spectrum gives a cutoff much too shallow to account for the sharp 
break seen in the data, e.g., $\chi^2_\nu =220/76$ for the data set 2. (The 
best fit parameters for this data set are: $\alpha=0.52$, $\ec=270$ keV, 
$\Omega/2\pi= 0.28$.) 

However, e-folded power laws approximate only roughly thermal Comptonisation 
spectra, as discussed by, e.g., Poutanen \& Svensson (1996; hereafter PS96). 
Therefore, we then use a thermal-Compton disc-corona model of PS96. In the 
model used, the hot coronal plasma is assumed to form hemispheres (with the 
radial Thomson optical depth of $\tau$) on the surface of an accretion disc. 
This model, however, is also very strongly ruled out by the data, yielding 
$\chi_\nu^2 \sim 1000/75$. The main reason for the bad fit is the presence of 
an {\it anisotropy break\/} in the underlying model spectrum. The anisotropy 
break appears in the disc-corona spectrum viewed from above (at $i=30^\circ$) 
around the peak energy of the second-order scattering because the first-order 
scattering is directed mainly towards the disc (Stern et al.\ 1995; PS96; 
Svensson 1996). For the seed blackbody photons at the temperature of 130 eV 
(Ba{\l}uci\'nska-Church et al.\ 1995) and the hot plasma temperature of $\sim 
100$ keV, the anisotropy break energy is $\sim 6$ keV. The \ginga\/ data are 
instead very well described by a {\it single\/} power law up to 30 keV and a 
Compton-reflection component on top of it, as found above. A secondary problem 
with the model of PS96 applied to the present data is the predicted solid 
angle of the reflector of $\Omega= 2\pi$, whereas the data prefer a lower 
value. Still, allowing $\Omega<2\pi$ in the model (which may correspond to a 
cold disc present only at a limited range of radii) does not result in an 
acceptable fit. E.g., the fit to the data set 1 (Fig.\ 2) yields $\chi_\nu^2 
=160/75$ for $kT=92$ keV, $\tau=2.2$, $\Omega/2\pi = 0.24$ for $i=30^\circ$. 
The model can be strongly ruled out ($\chi^2=117$) even for the largest 
viewing angle allowed by the optical data, $i=67^\circ$ (Dolan \& Tapia 1989)
The reason for the bad fit of the corona model is still the presence of the 
anisotropy break in the incident model spectrum. E.g., the incident spectrum 
in Figure 2 (dashed curve) can be approximated by a harder power law, 
$\alpha\simeq 0.43$, below 5.8 keV and a softer one, $\alpha\simeq 0.56$ above 
it. The pattern of residuals in Figure 2 clearly show how this broken 
power-law--like model does not fit the data (which are consistent with a {\it 
single\/} incident power law). 

We have also considered the homogeneous disc-corona model of Haardt \& 
Maraschi (1993) as a phenomenological description of the data (although 
that model does not satisfy the energy balance in the case of Cyg X-1, 
see Section 3.2). For that model, we use the code of Haardt (1993). The 
main difference with respect to the hemisphere model above is that now 
there is strong Comptonisation of the reflection component by the hot 
corona with $\tau\sim 1$ above the disc (see, e.g., PS96). Again, no 
acceptable fit is obtained, $\chi_\nu^2\sim 200/75$. Thus, the spectral 
data of Cyg X-1 {\it rule out\/} the disc-corona model regardless of the 
corona geometry. 

\begin{figure}
\begin{center}
\leavevmode
\epsfxsize=8.4cm \epsfbox{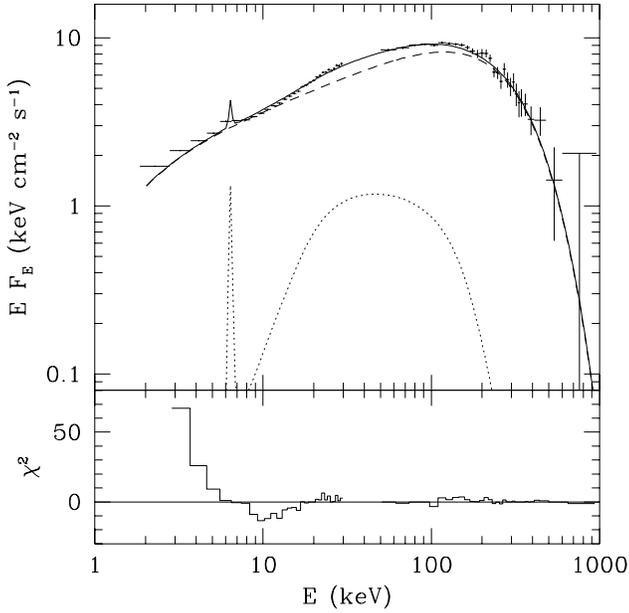}
\end{center}
\label{fig:break}
\caption{The anisotropic disc-corona model (PS96) fitted to the data set 1. 
The upper panel shows the data (crosses), the thermal Comptonisation continuum 
(dashed curve), and the reflection component (dotted curve). The solid curve 
shows the sum. The lower panel in this figure and in figures below shows the 
contribution to the total $\chi^2$ from separate data channels [multiplied by 
the sign of (data $-$ model)]. This model provides no satisfactory 
fit to the data due to the incident spectrum having a break at $\sim 5$ keV. } 
 \end{figure}

Motivated by our ruling out the disc-corona model, we have considered thermal 
Comptonisation in a hot plasma cloud irradiated approximately {\it 
isotropically\/} by soft seed photons. This removes the presence of the 
anisotropy break in the spectrum, but it retains the proper description of 
the high-energy cutoff from Comptonisation. To this end, we used an 
isotropic-scattering version of the Comptonisation code of PS96 (J. Poutanen, 
private communication). We have assumed that a fraction of 0.3 of the 
reflection spectrum undergoes further thermal Comptonisation in the hot cloud, 
which is consistent with source geometry inferred from energy balance (Section 
3.2 below); the fit results are insensitive to this number as long as it is 
$\ll 1$. Also, the fits are insensitive to the viewing angle; if $i$ is 
increased from our default value of $30^\circ$ to $67^\circ$, $\chi^2$ remains 
the same within $\pm 1$ for the four data sets. This model does not allow the 
reflector to be ionised, which may result in some overestimate of $\chi^2$ 
since the \ginga\/ data indicate the reflector is weakly ionised. The Fe 
abundance of this model consistent with all 4 data sets is $\af= 1$. As 
expected, we obtain a dramatic improvement of the fits, with $\chi_\nu^2 
\simless 1$ now. The fit results for the four data sets are given in Table 2, 
and the fit to the data set 2 is presented in Figure 3. We see that the hot 
plasma parameters are $kT\simeq 100$ keV and $\tau\simeq 2$ for all data sets. 

\begin{figure}
\begin{center}
\leavevmode
\epsfxsize=8.4cm \epsfbox{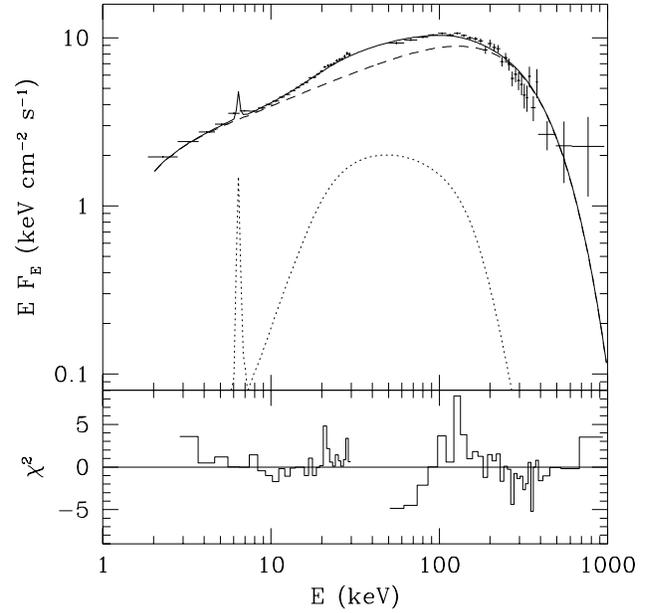}
\end{center}
\label{fig:ism}
\caption{The isotropic one-temperature thermal Comptonisation model fitted to 
the data set 2. The upper panel shows the data (crosses), the thermal 
Comptonisation continuum (dashed curve) and the reflection component (dotted 
curve). The solid curve shows the sum. This model yields $\chi_\nu^2<1$, but 
there is still a systematic residual pattern for the OSSE data (the lower 
panel).
 }\end{figure}

We have tested the isotropic Comptonisation code of PS96 by our own Monte 
Carlo calculations (see Zdziarski, Johnson \& Magdziarz 1996). We assumed a 
spherical source illuminated by seed photons at the center. We obtain the same 
plasma temperature, $kT\simeq 100$ keV, but the radial optical depth of the 
sphere is $\tau\simeq 1.3$. The difference between this $\tau$ and $\tau\simeq 
2$ obtained above is explained by the difference between the spherical source 
in the Monte Carlo calculations and the hemisphere source in the method of 
PS96. 

\begin{table*}
\label{t:fits1}
\centering
\caption
{The parameters of the fits with the single-temperature isotropic model. $I$ 
is the 1-keV normalisation in cm$^{- 2}$ s$^{-1}$, $kT$ is in keV, and EW is 
in eV. The column density, $\nh$ (in units of 10$^{22}$ cm$^{-2}$), is 
constrained to be $\geq$ the lower limit on the Galactic column. The errors 
are given for 90 per cent confidence intervals, $\Delta \chi^2=2.7$ (Lampton 
et al.\ 1976), and $\chi^2$ is given for 76 d.o.f.} 
\begin{tabular}{lccccccc}
\hline
Obs. &$\nh$ &$kT$ &$\tau$
     &$\Omega/2\pi$ &$I$
     &EW &$\chi^2$\\
1 &$0.4^{+0.14}_{-0}$ &$103^{+7}_{-5}$ &$1.98^{+0.09}_{-0.12}$
  &$0.34^{+0.03}_{-0.03}$ &1.19
  &$137^{+27}_{-29}$ &69\\
2 &$0.4^{+0.09}_{-0}$ &$99^{+6}_{-5}$ &$2.04^{+0.09}_{-0.11}$
  &$0.32^{+0.04}_{-0.03}$ &1.34
  &$137^{+29}_{-28}$ &97\\
3 &$1.19^{+0.18}_{-0.17}$ &$101^{+8}_{-8}$ &$2.08^{+0.12}_{-0.13}$
  &$0.19^{+0.03}_{-0.03}$ &0.79 
  &$121^{+30}_{-29}$ &60\\
4 &$0.78^{+0.20}_{-0.21}$ &$120^{+22}_{-16}$ &$1.74^{+0.24}_{-0.27}$
  &$0.25^{+0.04}_{-0.04}$ &0.74
  &$111^{+31}_{-32}$ &64\\
\hline
\end{tabular}
\end{table*}

\begin{table*}
\label{t:fits}
\centering
\caption
{The parameters of the fits with the two-component thermal-Comptonisation 
model. The parameters of the main component are denoted by the same symbols as 
in Table 1, and $I'$ and $kT'$ are the 1-keV normalisation and temperature of 
the secondary component in units of cm$^{- 2}$ s$^{-1}$ and keV, respectively. 
$L$ is the 2--1000 keV luminosity in $10^{37}$ erg s$^{-1}$ for the assumed 
distance of 2.5 kpc, and $\chi^2$ is given for 74 d.o.f.} 
\begin{tabular}{lcccccccccc}
\hline
Obs. &$\nh$ &$kT$ &$\tau$
     &$\Omega/2\pi$ &$I$ &$kT'$ &$I'$
     &EW &$L$ &$\chi^2$\\
1 &$0.85^{+0.26}_{-0.27}$ &$104^{+17}_{-17}$ &$1.76^{+0.32}_{-0.26}$
  &$0.49^{+0.07}_{-0.07}$ &1.33 &$51^{+9}_{-7}$ &0.0080
  &$115^{+31}_{-30}$ &3.7 &48\\
2 &$0.95^{+0.27}_{-0.27}$ &$107^{+19}_{-17}$ &$1.67^{+0.29}_{-0.28}$
  &$0.53^{+0.07}_{-0.07}$ &1.53 &$47^{+6}_{-5}$ &0.0106
  &$110^{+32}_{-33}$ &4.1 &64\\
3 &$1.50^{+0.29}_{-0.30}$ &$108^{+14}_{-15}$ &$1.87^{+0.17}_{-0.24}$
  &$0.25^{+0.06}_{-0.05}$ &0.83 &$43^{+21}_{-13}$ &0.0026 
  &$104^{+32}_{-28}$ &2.6 &56\\
4 &$1.18^{+0.28}_{-0.28}$ &$127^{+27}_{-23}$ &$1.53^{+0.30}_{-0.28}$
  &$0.32^{+0.08}_{-0.06}$ &0.79 &$36^{+13}_{-10}$ &0.0042 
  &$87^{+32}_{-35}$ &2.3 &54\\
\hline
\end{tabular}
\end{table*}

Although $\chi_\nu^2<1$ for the isotropic Comptonisation model, there are 
still strong systematic residuals in the fit to the OSSE data, see the lower 
panel of Figure 3. Specifically, the model spectrum cuts off at high energies 
less than the data, i.e., the model spectrum is broader than the observed 
spectrum. Such an effect could appear if there were an additional narrow 
spectral component at high energies. Such a sharp spectral component can be 
due to a hot, optically-thick plasma component in the source, e.g., a 
transition region between the hot and the cold medium in the accretion flow 
(Misra \& Melia 1996). 

We have therefore considered models with two hot-plasma components. The first 
component, along with its reflection, is computed in the same way as in the 
isotropic Comptonisation model above. The Fe abundance is kept at 
$\af=1.0$, which is both the average value from fits to the four spectra and a 
value within the confidence limits of all of the four fits. The second, 
additional component (whose reflection is not considered) is due to 
optically-thick thermal Comptonisation and it is computed using a modified 
Kompaneets equation as in Zdziarski et al.\ (1996). We see in Table 3 and 
Figure 4 that addition of a second Comptonisation component with $kT'\sim 50$ 
keV improves significantly the fit. In the actual fits, we have had to prevent 
the secondary component from giving a dominant contribution at low energies. 
This is achieved by fixing its low-energy spectral index to $\alpha'=0.15$, 
which we found to have a negligible effect on $\chi^2$. The Thomson optical 
depth of the second plasma component has been found by the Monte Carlo method. 
It is $\tau' \simeq 6$ for a sphere with a central source of seed photons. The 
luminosity in the best-fit optically-thick component is 13, 15, 5, and 8 per 
cent of the total luminosity for the data sets 1, 2, 3, and 4, respectively. 
This model provides excellent fits to the data ($\chi^2\sim 50$--$60/74$ 
d.o.f.), and the improvement with respect to the one-temperature model is 
statistically significant at more than 99.9 per cent (the probability that the 
reduction of $\chi^2$ is by chance is $\simeq 10^{-13}$) taking into account 
all four data sets. The two-component model strongly favours a face-on 
geometry; when $i$ is increased from $30^\circ$ (assumed above) to 
$i=67^\circ$, $\Delta\chi^2\simeq  +7$ for each of the data set 1 and 2. 
 
\begin{figure}
\begin{center}
\leavevmode
\epsfxsize=8.4cm \epsfbox{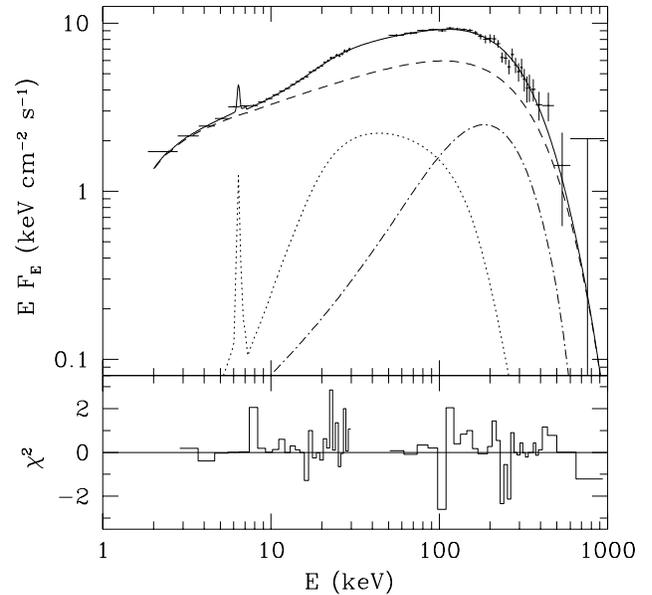}
\end{center}
\label{fig:2c}
\caption{The two-temperature thermal Comptonisation fit to the data set 1. The 
dashed and dot-dashed curves represent the primary and secondary Comptonisation 
components, respectively. The dotted curve represents Compton reflection 
of the primary component. The solid curve gives the sum. This model provides a 
statistically satisfactory description of the data.
}
 \end{figure} 

A possible alternative explanation to the hot plasma with a distribution of 
$kT$ and $\tau$ is Comptonisation by a plasma with nonthermal electron 
distribution. We have performed a number of Monte Carlo computations using 
either a Maxwellian electron distribution truncated at high energies or 
electron power-law distributions constrained to low values of the Lorentz 
factor ($\simless 2$). Our conclusion is that those models can in principle 
provide high-energy cutoffs stronger than those from the Maxwellian. However, 
$\chi^2$ fitting of Monte Carlo results would require a very extensive grid of 
models. Instead, an iterative-scattering method (PS96) modified for 
Comptonisation by a non-Maxwellian distribution appears to provide a simpler 
solution to this problem (J. Poutanen, private communication). We leave 
detailed investigation of this issue for future work. 

\subsection{Energy balance, geometry, and pair production}

In Section 3.1 above, we have ruled out the disc-corona model for Cyg X-1 
based on spectral fitting alone. This conclusion can be further supported 
by considering the energy balance between the hot and the cold phase.

The presence of Compton reflection shows there is cold matter in the vicinity 
of the X\g\ source. Only some of the incident X\g\ photons get reflected (with 
the integrated albedo of $\sim 0.15$, e.g., Magdziarz \& Zdziarski 1995), but 
most ($\sim 0.85$) are absorbed and reemitted in soft X-rays with a 
distribution close to a blackbody. Depending on the geometry, a fraction of 
the blackbody photons will return to the X\g\ source and provide some of the 
seeds for thermal Compton upscattering, which cools the hot plasma. (Further 
cooling can be provided by a blackbody flux from dissipation of gravitational 
energy in the cold matter). The ratio of the luminosity in the X\g\ spectrum 
to that of the seed photons (the Compton amplification factor) determines the 
resultant spectral shape because, at a given optical depth, the electron 
temperature is determined by the amount of cooling photons incident on the hot 
plasma. From Monte-Carlo simulations, we find that the rather flat continuum 
spectrum ($\alpha\sim 0.6$) extending out to $\sim 150$ keV can only be 
produced if the incident seed photon luminosity is about 14 times less than 
the luminosity in the Comptonised X\g\ spectrum.  This rules out a homogeneous 
disc-corona geometry of the source, in which approximately half of the hard 
X-ray photons are intercepted by the disc, 85 per cent of which are 
thermalised, and then re-intercepted by the corona, giving an X-ray luminosity 
only $\sim 2.4$ times that of the seed photons.  The lack of soft photons 
could in principle be obtained if the hot corona is in the form of detached 
active regions (e.g., PS96). However, this would imply an observed solid angle 
covered by the reflecting medium as seen from the hot source close to $2\pi$ 
whereas the observed value is $\sim 0.4 \times 2\pi$. This means 
that the hot source does {\it not\/} form a corona (either homogeneous or 
patchy) above the surface of a cold disc, fully consistent with the results of 
spectral fitting in Section 3.1. 

Assuming that the reflection spectrum is not strongly Comptonised, i.e.,
the observed amount of reflection is indicative of the solid angle 
subtended by the cold material to the hard X-ray source, then we can 
estimate the fraction, $g$, of the reprocessed radiation that returns to 
the hot source to form the Compton seed photons. From energy balance 
considerations  $g \le (0.85 \times 0.5 \times \Omega/2\pi \times 14)^{-1} 
\simeq (6 \times \Omega/2\pi)^{-1} \sim 0.4$. (A similar value was used in the 
fits in Section 3.1.) Again this highlights the problems of the homogeneous 
corona-disc geometry, in which all the reprocessed photons are re-intercepted 
by the hot plasma. 

The considerations above explain the fit results in which we have been 
unable to obtain satisfactory models with the disc-corona model (see 
above). A possible source geometry consistent with the observed spectra 
is a hot inner region together with a colder outer disc (e.g.\ Shapiro, 
Lightman \& Eardley 1976; Bj\"ornsson et al.\ 1996; Narayan 1996). This 
geometry explains the observed low reflection fraction, the large 
Compton amplification of seed photons, as well as the narrowness of the 
K$\alpha$ line observed by E96. Also, the seed soft photons in this model are 
not produced underneath the hot plasma but rather come from various directions 
sideways. This then explains the absence of the anisotropy break in the 
spectrum. We intend to further investigate the expected spectrum from such 
geometry in future work. 

The last issue we consider is the presence of \ee\ pairs in the hot source. 
Knowing the shape of the spectrum and $\tau$ ($\sim 1$--2), we can compute the 
compactness parameter, $\ell\equiv L\sigma_{\rm T}/R m_{\rm e} c^3$ (where $R$ 
is the source size), at which photon-photon pair production rate is sufficient 
to produce pairs with the given $\tau$ (Svensson 1984). We obtain $\ell\sim 
40$--70 roughly correlated with the luminosity varying in the range of $\sim 
0.02$--0.03 of the Eddington limit (assuming the mass of $10M_\odot$). This 
implies an approximately constant lower limit on the source size of $R\sim 5$ 
Schwarzschild radii. The source is pair-dominated at this limit, and is 
electron-dominated if $R$ is much larger. The constancy with $L$ is due to an 
approximate constancy of the model fluxes seen around 0.5 MeV (in spite of the 
overall X\g\ flux varying), see Figure 1. 

We also note that the pair annihilation feature at the $kT$ and $\tau$ implied 
by the thermal Comptonisation model with \ee\ pairs would be rather weak, with 
the peak annihilation flux less than 0.1 of the observed 0.5 MeV flux (see 
Macio{\l}ek-Nied\'zwiecki, Zdziarski \& Coppi 1995), and the luminosity in 
annihilation photons of less than $\sim 2$ per cent of the total luminosity. 
Thus, the observed spectra are consistent with (but do not require) the hot 
($kT\sim 100$ keV) source made of \ee\ pairs.   

Note that virtually all pair production is due to the hotter and optically 
thinner plasma component, with $kT\simeq 100$ keV and $\tau\sim 1$--2. This 
can be seen in Figure 4, in which number of Wien photons around 0.5 MeV is 
much less than that in the main spectral component. Consequently, the cooler, 
optically thick, plasma ($kT\simeq 50$ keV, $\tau\sim 6$) is made entirely of 
electrons and protons. 

\section{CONCLUSIONS}

We find the X\g\ source in Cyg X-1 does not form a corona above the surface of 
an accretion disc. Three arguments lead us to that conclusion: the solid angle 
subtended by the cold medium as seen from the X\g\ source is significantly 
smaller than $2\pi$, the X\g\ source is photon-starved, and there is no 
anisotropy break (necessary in the disc-corona models) in the X-ray spectrum. 
In addition, the reflecting cold medium is far away from the central black 
hole, as demonstrated by the narrowness of the fluorescent K$\alpha$ line 
observed by \asca\/ (E96). All the properties above are consistent with a hot 
inner disc Comptonising soft photons from a cold outer disc. This 
distinguishes Cyg X-1 from some Seyfert 1s, in which the cold, reflecting, 
disc appears to extend all the way to the innermost stable orbit (e.g., Fabian 
et al.\ 1995). Note that this difference cannot be explained by the 
radiation-pressure disc instability in a disc-corona system, which is stronger 
for more massive black holes (e.g., Svensson \& Zdziarski 1994). 

The most likely radiative process producing X\g\ photons in Cyg X-1 is 
Comptonisation of seed soft X-rays photons. The observed sharpness of 
the high-energy cutoff rules out a uniform thermal plasma. Either a 
distribution of plasma parameters or an electron distribution different 
from a Maxwellian is required. A remarkable property of the spectra is 
their almost constant shape with the varying amplitude. In models with 
repeated Compton scattering, this implies the constancy of the plasma 
parameters (e.g., $\tau$ and $kT$ for thermal Comptonisation) with 
changing accretion rate. This represent a strong constraint on future 
models of Cyg X-1. In particular, {\it simplest\/} thermal \ee\ pair 
models predict a strong dependence of $\tau$ on the luminosity (e.g.\ 
Zdziarski 1985). The constant spectral shape can be achieved in models 
with high-energy nonthermal electrons (e.g., Svensson 1987; Lightman \& 
Zdziarski 1987). However, these models predict a strong annihilation feature 
which is not seen in the data (P96), although this constraint could be 
possibly weakened if the annihilation rate were reduced by having the pairs 
reaccelerated before they lose energy completely. 

\section*{ACKNOWLEDGEMENTS}

This research has been supported in part by the Polish KBN grants 2P03D01008 
and 2P03D01410 and NASA grants and contracts. We are grateful to Juri Poutanen 
for providing us with his disc-corona code and extensive discussions. We also 
thank Pawe\l\ Magdziarz for his assistance with implementing models into the 
{\sc xspec} software package.

\end{document}